\documentclass[12pt]{iopart}

\usepackage{graphicx}
\usepackage[american]{circuitikz}
\usetikzlibrary{arrows.meta,fit}
\usetikzlibrary{decorations.markings}
\usetikzlibrary{positioning}
\usepackage{tikz}
\usepackage{hyperref}
\usepackage{tabularx}

\newcommand{\st}[1]{_\mathrm{#1}} 

\usepackage[none]{hyphenat}
\definecolor{dark_red}{RGB}{190,1,1}
\definecolor{dark_green}{RGB}{1,164,1}

\begin{document}
	
	\title[Energy Shift with Coupling (ESC): a new quench protection method]{Energy Shift with Coupling (ESC): \\ a new quench protection method}
	
	\author{E.~Ravaioli, A.~Verweij, M.~Wozniak}
	
	\address{CERN, Meyrin, Switzerland}
	\ead{Emmanuele.Ravaioli@cern.ch}
	\vspace{10pt}
	\begin{indented}
		\item[]August 2024
	\end{indented}
	
	\begin{abstract}
		Quench protection of full-size, high-field accelerator magnets poses significant challenges. Maintaining the hot-spot temperature and peak voltage-to-ground within acceptable limits requires a protection system that quickly transitions most of the coil turns to the normal state.
		Existing magnet protection technologies, such as quench protection heaters or the Coupling Loss Induced Quench system (CLIQ), have been successfully applied. However, they both present shortcomings since they require either thin insulation between the heaters and the magnet conductor – in the case of heaters, or direct electrical connections to the magnet coil – in the case of CLIQ.
		A novel quench protection method, Energy Shift with Coupling (ESC), is presented, which can achieve excellent quench protection performance without requiring any electrical connection or close physical contact with the coils to protect.
		ESC relies on normal-conducting auxiliary coils strongly magnetically coupled with the magnet coils to protect. Upon quench detection, capacitive units connected across the auxiliary coils introduce a high current change in the auxiliary coils, causing a rapid shift of magnet stored energy from the magnet coils to the auxiliary coils. This has three beneficial effects: sudden reduction of ohmic loss in the normal zone of the magnet conductor, introduction of high transient losses in the magnet conductor, thus causing a quick transition to the normal state, and extraction of a part of the magnet's stored energy to the auxiliary coils. 
		The applicability of the ESC concept on an existing magnet design is analyzed with electro-magnetic and thermal transient simulations performed with the STEAM-LEDET program. The advantages and disadvantages of ESC are discussed and compared to other conventional quench protection methods.
		Simulation results show that ESC can be applied to protect full-scale magnets with reasonable requirements in terms of size and location of the auxiliary coils, and of capacitive unit parameters.
	\end{abstract}
	
	%
	\vspace{2pc}
	\noindent{\it Keywords}: accelerator magnet, coupling loss, multiphysics modeling, quench protection, simulation
	%
	%
	%
	%

	\section{Introduction}
	The pursuit of higher magnetic fields in very compact superconducting magnets results in designs characterized by higher energy density and current density.
	It is mandatory to consider during the design phase a strategy to mitigate the possibly destructive consequences of a quench, which is an unexpected transition to the normal state occurring in a spot of the superconductor.
	Reducing below safe values the peak temperature, voltage to ground, and stress of the conductor during a quench transient is a significant challenge, and calls for a fast and effective active quench protection system that can detect the quench and either extract part of the magnet's stored energy, or distribute it more uniformly in the coil winding pack~\cite{Maddock_Protection,1018662,Sonnemann:499591,Bottura_Quench101,thesis_CLIQ}.
	Furthermore, for many applications it is critical to design a quench protection system that is robust, redundant, and effective for full-size magnets.
	The protection system should be as simple, small, and inexpensive as possible, easy to integrate in the magnet circuit, and not interfering with other systems.
	For the main bending and focussing/defocussing accelerator magnets, it is also important to retain the ability to protect individual magnets that are part of a circuit containing many magnets powered in series~\cite{305593,792249,MT24_Chain}.
	
	Conventional quench protection methods such as energy-extraction (EE)~\cite{614456,Hagedorn:1996uy,Schmidt:466521,988140}, quench protection heaters (QH)~\cite{Rodríguez-Mateos:692002,Rodríguez-Mateos:466520,988141,Salmi:2132789,Szeless:615758,5067266,Eucas2107_QH_MQXFS3,8280523}, and the Coupling-Loss Quench Protection system (CLIQ)~\cite{CLIQ_MT23,thesis_CLIQ} each come with important limitations.
	In fact, EE by principle requires imposing a significant voltage across the protected magnet.
	This makes it difficult to scale to large, high current-density magnets, and not suitable for protecting high-field magnets that are powered in series, since its effectiveness is strongly affected by the total circuit inductance~\cite{Bottura_Quench101}.
	
	QH are widely used to protect high-field low-temperature superconductor (LTS) magnets due to their easiness to scale to longer length and to add redundancy by installing separate, independent protection units.
	QH rely on thermal diffusion to heat up parts of the coil winding pack until they are transitioned to the normal state.
	As~a result, the magnet stored energy is distributed more uniformly in the coil turns, and the magnet current is discharged more quickly due to the developed magnet coil resistance.
	Their main limitations are the challenge to achieve quick and simultaneous heating of the entire coil volume, calling for thin insulation thickness and thoroughly distributed heater strips, and to achieve reliable high electrical robustness, calling for thick insulation thickness.
	
	CLIQ offers a valid alternative that achieves highly effective power deposition relying on transient losses, primarily inter-filament coupling loss~\cite{Morgan70,Carr74,thesis_Arjan}, and that utilizes electrically-robust current leads.
	It was successfully tested on magnets of different geometries, superconductor types, and sizes~\cite{CLIQ_Cryogenics,CLIQ_SuST,CLIQ_ASC2014,EUCAS2015_CLIQonMQY,MT24_CLIQonMB}.
	CLIQ's main disadvantages are the direct connection to the magnet coils, which is required by principle and could be problematic when applied to a circuit including many magnets connected in series, the cryogenic load associated to the additional current leads, the relatively high voltages induced directly in the magnet coils, the difficulty of achieving redundancy, and non-standard forces introduced in the magnet structure.
	
	Other quench protection systems have been proposed.
	The possibility of heating up the superconductor using well-insulated resistive coils was explored in different forms.
	Small coils with very low mutual coupling to the magnet to protect were powered with an AC current to serve as inductive heating stations~\cite{Bromberg_2011,6072242,Agustsson_2013,Bromberg_2006,10076805}.
	Furthermore, larger scale coils with very good magnetic coupling were used to induce transient loss in the superconductor~\cite{CLIQ_Cryogenics,thesis_CLIQ}.
	These systems have the advantage of not relying on thermal diffusion, and hence allowing the use of thick and reliable insulation.
	
	The effectiveness of coupled secondary circuits formed by well electrically conducting elements in the magnet structure is well known in the literature~\cite{osti_5334172,osti_7278286,1060158}.
	The introduction of coupled coils to absorb part of the magnet's stored energy was proposed~\cite{7360114,Mentink_2017}.
	This method requires careful integration in the magnet design, and can effectively reduce the hot-spot temperature by reducing the energy deposited in the coil volume.
	However, these methods require by principle a driving voltage to transfer the energy to the coupled secondary elements, either from an energy-extraction system or from the resistive voltage build-up in the magnet coil, which needs to be heated up by a different system.
	
	The innovative concept of coupled coils with the dual function of quickly introducing high coupling loss, hence rapidly transitioning the magnet coils to the normal state, and of extracting energy due to magnetic coupling, was introduced with the Secondary-CLIQ (S-CLIQ) method~\cite{Mentink_2020,10330023}.
	With respect to CLIQ, this solution does not have any direct electrical connection to the magnet coils, does not induce high voltages across the magnet turns, and can more easily be made redundant.
	Its main disadvantages are the introduction of auxiliary coils in the magnet design and the asymmetric mechanical forces arising in the magnet and auxiliary coils.
	When S-CLIQ is activated, virtually no change in the magnet transport current is induced until the magnet coil turns resistive.
	
	In this contribution, a novel quench protection method, from now on named \textbf{Energy Shift with Coupling} (\textbf{ESC}), is presented, which offers multiple advantages in terms of effectiveness, robustness, reliability, and easiness of implementation.
	The method consists in introducing a current pulse through normal-conducting coils that are insulated from the magnet coils and are strongly magnetically coupled to them.
	This has a threefold effect on the magnet protection.
	First, an almost immediate reduction of the magnet transport current is induced, which decreases the ohmic loss and magneto-resistance in the magnet coil's hot-spot where the quench started.
	Second, a very high magnetic-field change, in the order of hundreds of T/s and maintained for at least a few milliseconds, is introduced, which aims at quickly transitioning the entire magnet coil volume to the normal state due to transient losses.
	Finally, part of the magnet's stored energy is transferred from the magnet coils to the normal-conducting ESC coils due to magnetic coupling.
	
	While the effort to experimentally demonstrate the ESC technology is ongoing, this contribution focusses on the concept introduction.
	The ESC electrical scheme and its working principle are presented in detail.
	Characteristic electro-magnetic and thermal transients occurring in a magnet during an ESC discharge are analyzed.
	ESC performance is compared to that of conventional quench protection systems.
	More ambitiously, its role in enabling the protection of future high magnetic-field magnets even in a new design space is highlighted.

	\section{Energy Shift with Coupling concept}
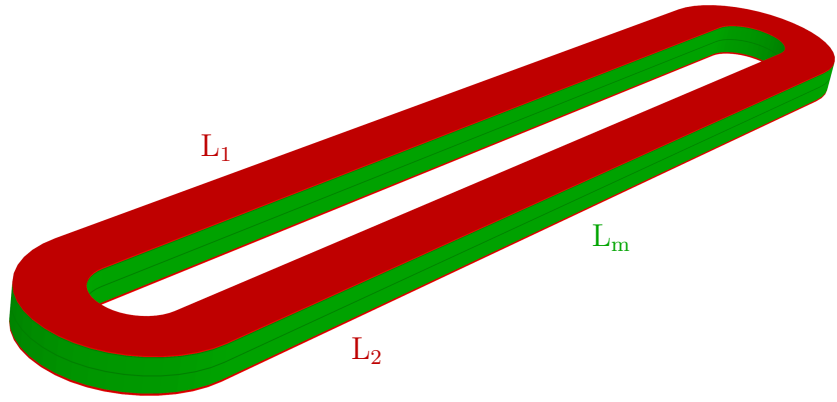
\begin{figure}[!t]
		\centering
		\includegraphics[width=0.7\textwidth]{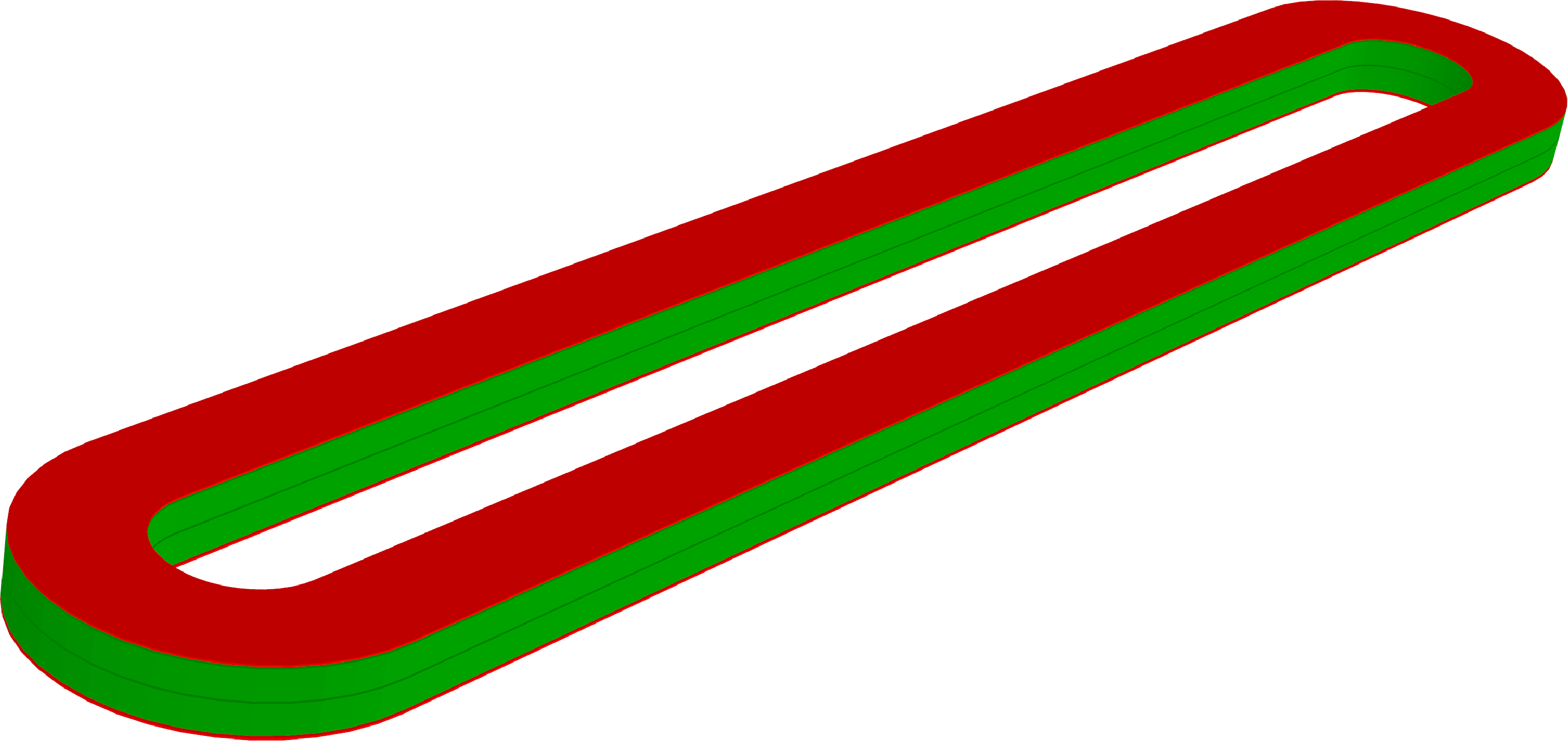}
		\begin{tikzpicture}[overlay, remember picture]
		\node[dark_green, anchor=east, align=center] at (-2.7cm,+2.1cm) {L$\st{m}$};
		\node[dark_red, anchor=east, align=center] at (-8.0cm,+3.3cm) {L$\st{1}$};
		\node[dark_red, anchor=east, align=center] at (-6.0cm,+0.6cm) {L$\st{2}$};
		\end{tikzpicture}
		\caption{Three-dimensional sketch of a superconducting racetrack-coil magnet (L$\st{m}$, in green) protected by two ESC resistive coils (L$_1$ and L$_2$, in red) located at the top and bottom of the magnet coils.
			The current leads at the coil ends are not displayed.
			\label{fig_sketch_ESC}
		}
	\end{figure}

	ESC can be applied to protect magnets of different geometries.
	In this publication, the example of a racetrack coil magnet protected by two ESC coils, whose three-dimensional sketch is shown in Figure~\ref{fig_sketch_ESC}, is taken.
	The ESC coils are resistive and galvanically insulated from the magnet coils.
	Due to their proximity and similar geometry, they are strongly magnetically coupled to the magnet coils, and they are also mutually coupled to each other.
	When a voltage $U\st{c1}$~[V] is applied across an ESC coil, its current $I_1$~[A] is changed, and hence an inductive voltage is developed across the magnet coil, which in turn imposes a change of the magnet current $I\st{m}$~[A].
	This allows introducing a high magnet current change without imposing a high voltage across the magnet coil, which is a major advantage of the ESC method.
	In fact, consider the interaction between two coupled inductors in closed loops, with self-inductance $L\st{m}$~[H] and $L_1$~[H], and mutual inductance $M\st{m,1}$~[H], which is described by the following set of equations:
	\begin{equation}
	\label{eq_00a}
	\fl
	\cases{
		L\st{m} \frac{dI\st{m}}{dt} + M\st{m,1} \frac{dI\st{1}}{dt} = 0 \\
		L\st{1} \frac{dI\st{1}}{dt} + M\st{m,1} \frac{dI\st{m}}{dt} + U\st{c1} = 0 \\
	}
	\end{equation}
	It is easily shown that the magnet coil current change~is
	\begin{eqnarray}
	\label{eq_00b}
	\fl \frac{dI\st{m}}{dt}
	= \frac{U\st{c1}}{L_1} \frac{M\st{m,1}}{L\st{m}} \frac{1}{1 - k\st{m,1}^2}
	= \frac{U\st{c1}}{\sqrt{L\st{m} L_1}} \frac{k\st{m,1}}{1 - k\st{m,1}^2} ,
	\end{eqnarray}
	with $k\st{m,1}$$=$$M\st{m,1}/\sqrt{L\st{m}L\st{1}}$.
	This result demonstrates that it is possible to introduce a negative magnet coil current change without imposing any voltage across~L$\st{m}$, which in this example has its terminals shorted, by applying across the ESC coil a voltage $U\st{c1}$ that has the opposite sign to~$k\st{m,1}$.

	\subsection{ESC electrical scheme}
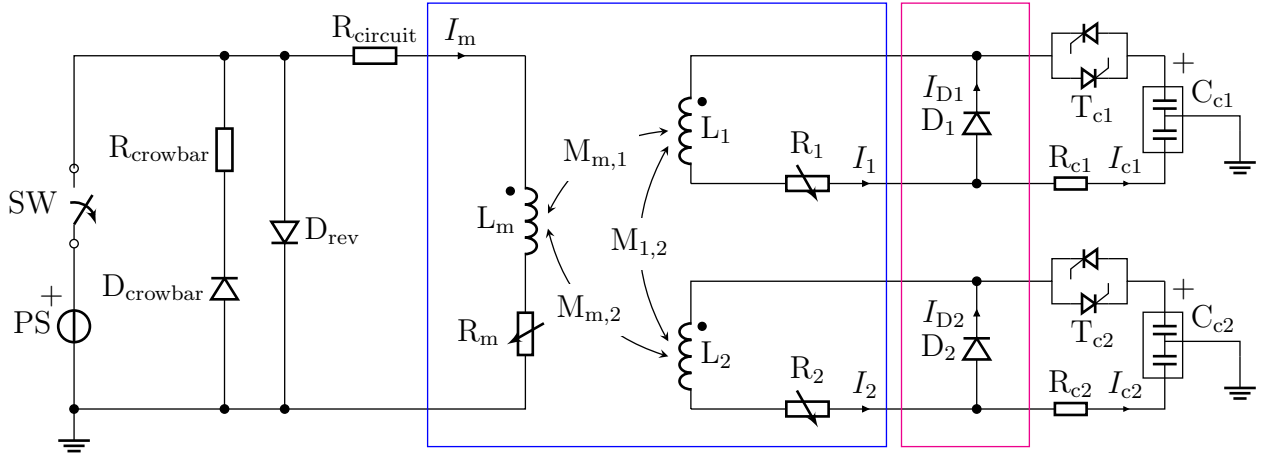
\begin{figure}[tp]
	\centering
	\begin{circuitikz}[line width=0.5pt]
		\centering
		
		\ctikzset{resistor = european}
		
		
		\coordinate (G1) at ( 0.0, 0.0-1.2);
		\coordinate (A1) at ( 0.0, 0.5-1.2);
		\coordinate (A2) at ( 0.0, 1.5-1.0);
		\coordinate (A3) at ( 0.0, 2.0-1.0);
		\coordinate (A4) at ( 0.0, 3.0-1.0);
		\coordinate (A5) at ( 0.0, 3.5);
		\coordinate (B1) at ( 2.0, -1.2);
		\coordinate (B2) at ( 2.0, -0.2);
		\coordinate (B3) at ( 2.0, 1.0);
		\coordinate (B4) at ( 2.0, 2.0-0.5);
		\coordinate (B5) at ( 2.0, 3.0);
		\coordinate (B6) at ( 2.0, 3.5);
		\coordinate (C1) at ( 6.0, 0.0-1.2);
		\coordinate (C2) at ( 6.0, 0.5-1.2);
		\coordinate (C3) at ( 6.0, 1.5-1.2);
		\coordinate (C4) at ( 6.0, 2.0-1.2);
		\coordinate (C5) at ( 6.0, 3.0-1.2);
		\coordinate (C6) at ( 6.0, 3.5);
		\coordinate (D4) at ( 8.2, 1.8);
		\coordinate (D5) at ( 8.2, 2.0);
		\coordinate (D6) at ( 8.2, 3.0);
		\coordinate (D7) at ( 8.2, 3.5);
		\coordinate (E2) at ( 8.5, 1.8);
		\coordinate (E3) at (11.0, 1.8);
		\coordinate (E4) at (12.0, 1.8);
		\coordinate (E5) at (12.0, 1.9);
		\coordinate (E6) at (12.0, 3.4);
		\coordinate (E7) at (12.0, 3.5);
		\coordinate (F4) at (14.5, 1.8);
		\coordinate (F8) at (14.5, 2.2);
		\coordinate (F5) at (14.5, 2.7);
		\coordinate (F6) at (14.5, 3.0);
		\coordinate (F7) at (14.5, 3.5);
		\coordinate (L1) at (14.35, 2.35);
		\coordinate (L2) at (14.6, 2.95);
		\coordinate (L3) at (14.5, 3.0);
		\coordinate (T1) at (14.0, 3.5);
		\coordinate (T2) at (14.0, 3.8);
		\coordinate (T3) at (13.0, 3.8);
		\coordinate (T4) at (13.0, 3.5);
		\coordinate (T5) at (13.0, 3.2);
		\coordinate (T6) at (14.0, 3.2);
		\coordinate (T7) at (13.1, 2.8);
		\coordinate (G4) at (15.5, 2.7);
		\coordinate (G5) at (15.5, 2.5);
		\coordinate (H1) at (14.75, 3.4);
		\coordinate (PS1) at (-0.3, 0.3);
		\coordinate (R1) at ( 2.8, 3.5);
		\coordinate (R2) at ( 2.8, -1.2);
		\coordinate[below=3cm of D4]  (D4b);
		\coordinate[below=3cm of D5]  (D5b);
		\coordinate[below=3cm of D6]  (D6b);
		\coordinate[below=3cm of D7]  (D7b);
		\coordinate[below=3cm of E2]  (E2b);
		\coordinate[below=3cm of E3]  (E3b);
		\coordinate[below=3cm of E4]  (E4b);
		\coordinate[below=3cm of E5]  (E5b);
		\coordinate[below=3cm of E6]  (E6b);
		\coordinate[below=3cm of E7]  (E7b);
		\coordinate[below=3cm of F4]  (F4b);
		\coordinate[below=3cm of F5]  (F5b);
		\coordinate[below=3cm of F6]  (F6b);
		\coordinate[below=3cm of F7]  (F7b);
		\coordinate[below=3cm of F8]  (F8b);
		\coordinate[below=3cm of L1]  (L1b);
		\coordinate[below=3cm of L2]  (L2b);
		\coordinate[below=3cm of L3]  (L3b);
		\coordinate[below=3cm of T1]  (T1b);
		\coordinate[below=3cm of T2]  (T2b);
		\coordinate[below=3cm of T3]  (T3b);
		\coordinate[below=3cm of T4]  (T4b);
		\coordinate[below=3cm of T5]  (T5b);
		\coordinate[below=3cm of T6]  (T6b);
		\coordinate[below=3cm of T7]  (T7b);
		\coordinate[below=3cm of G4]  (G4b);
		\coordinate[below=3cm of G5]  (G5b);
		\coordinate[below=3cm of H1]  (H1b);
		\coordinate[below=3cm of PS1]  (PS1b);
		\coordinate (M1) at ( 6.3, 1.4);
		\coordinate (M2) at ( 6.3, 1.2);
		\coordinate (M3) at ( 7.9, 2.3);
		\coordinate (P1) at ( 5.8, 1.7);
		\coordinate (P2) at ( 8.35, 2.89);
		\coordinate (P3) at ( 8.35, -0.1);
		\coordinate (Z1) at ( 8.5, 0.6);
		\coordinate (Z2) at ( 8.5, 1.2);
		\coordinate (Z3) at ( 8.5, 1.9);
		\coordinate (Z4) at ( 8.5, 2.5);
		\coordinate (Z5) at (12.0, 2.5);
		\coordinate (Z6) at (12.0, 0.6);
		\coordinate (Z7) at (10.3, 2.7);

		\draw[color=black] (G1) to (A1)
		to [/tikz/circuitikz/bipoles/length=20pt, european voltage source, l=PS] (A2)
		to (A3)
		to[opening switch, o-o, l=SW, mirror] (A4);
		\node[] at (PS1) {+};
		\draw[color=black] (A4) to (A5)
		to (B6)
		to [/tikz/circuitikz/bipoles/length=20pt, R, l=R$\st{circuit}$, i=$I\st{m}$] (C6);
		\draw[color=black] (G1) to (B1)
		to (B2)
		to [/tikz/circuitikz/bipoles/length=20pt, D, l=D$\st{crowbar}$] (B3)
		to (B4)
		to [/tikz/circuitikz/bipoles/length=20pt, R, l=R$\st{crowbar}$] (B5)
		to (B6);
		\draw[color=black] (B1) to (C1)
		to (C2)
		to [/tikz/circuitikz/bipoles/length=20pt, vR, invert, l=R$\st{m}$] (C3)
		to (C4)
		to [/tikz/circuitikz/bipoles/length=30pt, L, mirror, l=L$\st{m}$] (C5)
		to (C6);
		\draw[color=black] (R1)
		to [/tikz/circuitikz/bipoles/length=20pt, D, l=D$\st{rev}$] (R2);
		\draw[color=black] (G1) node[ground]{};
		
		\draw[color=black] (D4)
		to (E2)
		to [/tikz/circuitikz/bipoles/length=20pt, vR, mirror, l=R$\st{1}$, i=$I\st{1}$] (E3)
		to (E4)
		to [/tikz/circuitikz/bipoles/length=20pt, D, l=D$\st{1}$, i=$I\st{D1}$] (E6)
		to (E7)
		to (D7)
		to (D6)
		to [/tikz/circuitikz/bipoles/length=30pt, L, mirror, l=L$\st{1}$] (D5)
		to (D4);
		\draw[color=black] (E4) to [/tikz/circuitikz/bipoles/length=15pt, R, l=R$\st{c1}$, i=$I\st{c1}$] (F4)
		to (F8)
		to [/tikz/circuitikz/bipoles/length=15pt, C] (F5)
		to [/tikz/circuitikz/bipoles/length=15pt, C] (F6)
		to (F7);
		\node[draw, rectangle, fit={(L1) (L2)}] {};
		\node at (L1) [right=2mm of L3] {C$\st{c1}$}; 
		\draw[color=black]  (F7)
		to (T1)
		to (T2)
		to [/tikz/circuitikz/bipoles/length=0.5cm,thyristor] (T3)
		to (T4)
		to (E7);
		\draw[color=black] 
		(T4)
		to (T5)
		to [/tikz/circuitikz/bipoles/length=0.5cm, thyristor] (T6)
		to (T1);
		\node at (T7) [right=0mm of T7] {T$\st{c1}$}; 

		\draw[color=black] (F5) to (G4) to (G5);
		\draw[color=black] (G5) node[ground]{};
		\node[] at (H1) {+};
		
		\draw[color=black] (D4b)
		to (E2b)
		to [/tikz/circuitikz/bipoles/length=20pt, vR, mirror, l=R$\st{2}$, i=$I\st{2}$] (E3b)
		to (E4b)
		to [/tikz/circuitikz/bipoles/length=20pt, D, l=D$\st{2}$, i=$I\st{D2}$] (E6b)
		to (E7b)
		to (D7b)
		to (D6b)
		to [/tikz/circuitikz/bipoles/length=30pt, L, mirror, l=L$\st{2}$] (D5b)
		to (D4b);
		\draw[color=black] (E4b) to [/tikz/circuitikz/bipoles/length=15pt, R, l=R$\st{c2}$, i=$I\st{c2}$] (F4b)
		to (F8b)
		to [/tikz/circuitikz/bipoles/length=15pt, C] (F5b)
		to [/tikz/circuitikz/bipoles/length=15pt, C] (F6b)
		to (F7b);
		\node[draw, rectangle, fit={(L1b) (L2b)}] {};
		\node at (L1b) [right=2mm of L3b] {C$\st{c2}$}; 
		\draw[color=black] (F5b) to (G4b) to (G5b);
		\draw[color=black] (G5b) node[ground]{};
		\node[] at (H1b) {+};
		\draw[color=black]  (F7b)
		to (T1b)
		to (T2b)
		to [/tikz/circuitikz/bipoles/length=0.5cm,thyristor] (T3b)
		to (T4b)
		to (E7b);
		\draw[color=black] 
		(T4b)
		to (T5b)
		to [/tikz/circuitikz/bipoles/length=0.5cm, thyristor] (T6b)
		to (T1b);
		\node at (T7b) [right=0mm of T7b] {T$\st{c2}$}; 

		\draw [<->,>=stealth] (M1)  to [bend left]  node[pos=0.5,fill=white] {M$\st{m,1}$} ++(1.6,+1.1);
		\draw [<->,>=stealth] (M2)  to [bend right] node[pos=0.5,fill=white] {M$\st{m,2}$} ++(1.6,-1.7);
		\draw [<->,>=stealth] (M3)  to [bend right] node[pos=0.5,fill=white] {M$\st{1,2}$} ++(0.0,-2.6);
		\node[circle,draw=black,fill=black,minimum size=3pt,inner sep=0pt] at (P1) {};
		\node[circle,draw=black,fill=black,minimum size=3pt,inner sep=0pt] at (P2) {};
		\node[circle,draw=black,fill=black,minimum size=3pt,inner sep=0pt] at (P3) {};

		\node[circ] at (G1) {};
		\node[circ] at (B1) {};
		\node[circ] at (B6) {};
		\node[circ] at (R1) {};
		\node[circ] at (R2) {};
		\node[circ] at (E4) {};
		\node[circ] at (E7) {};
		\node[circ] at (E4b) {};
		\node[circ] at (E7b) {};
		
%
		\draw[blue] (4.7,-1.7) rectangle (10.8,4.2);
		\draw[magenta] (11.0,-1.7) rectangle (12.7,4.2);

%


	\end{circuitikz}

	\caption{Simplified electrical scheme of the circuit including a magnet protected by two ESC coils.
		The magnet circuit includes a power supply (PS), a switch (SW) that opens when PS is switched off, a crowbar composed of a resistor R$\st{crowbar}$ and a diode D$\st{crowbar}$, a reverse diode D$\st{rev}$ to protect the PS, warm circuit resistance R$\st{circuit}$, and a magnet with its inductance L$\st{m}$ and resistance R$\st{m}$.
		Each ESC circuit includes: auxiliary coils with their inductances (L$_1$, L$_2$) and resistances (R$_1$, R$_2$); an ESC unit composed of a capacitor bank grounded in its mid-point (C$\st{c1}$, C$\st{c2}$), an internal resistance (R$\st{c1}$, R$\st{c2}$) and back-to-back triggering thyristors (T$\st{c1}$, T$\st{c2}$); and a diode (D$_1$, D$_2$).
		The coils $L\st{m}$, $L_1$, and $L_2$ are mutually coupled.
		The blue rectangle identifies the parts of the circuit at cryogenic temperature.
		The ESC diodes (magenta rectangle) can be located either in the room temperature or cryogenic part of the circuit.
		}
	\label{fig_circuit}
\end{figure}

	The electrical scheme of the magnet circuit protected by two ESC circuits is shown in Figure~\ref{fig_circuit}.
	At the moment of the quench detection, the power supply PS is switched off to let the magnet current flow through its crowbar (R$\st{crowbar}$, D$\st{crowbar}$), and both ESC units are triggered simultaneously, hence introducing high positive current changes  in the ESC coils $dI_1/dt$ and  $dI_2/dt$~[A/s].
	As~a result, a negative current change $dI\st{m}/dt$~[A/s] is imposed in the magnet coil.
	The~diode D$\st{rev}$ is installed across the magnet to allow a path for a negative magnet circuit current in the case the ESC units are activated when the magnet initial current is very low.
\newcommand{\labelednode}[3][above]{
	\node[circle,draw,fill=purple!50,minimum size=2mm] (#2) at (#3) {};
	\node[#1,black] at (#2.west) {#2}; 
}

	Each ESC unit is composed of a charged capacitor bank (C$\st{c1}$, C$\st{c2}$) and a triggering system (T$\st{c1}$, T$\st{c2}$).
	The units are electrically insulated from the magnet and are individually grounded at the mid-point of their capacitor banks to halve the peak voltage to ground in the ESC coils after triggering.
	The ESC diodes D$_1$ and D$_2$ bring several operational advantages, which will be discussed in more detail in Section~\ref{sec_D_ESC}.
	
	The total magnetic energy stored in the magnet and two ESC coils~is
	\begin{eqnarray}
	\label{eq_00c}
	\fl E\st{mag}
	= \frac{1}{2} \left( L\st{m} I\st{m}^2 + L_1 I_1^2 + L_2 I_2^2 \right) + M\st{m,1} I\st{m} I_1 + M\st{m,2} I\st{m} I_2 + M_{1,2} I_1 I_2 .
	\end{eqnarray}
	Before triggering ESC, $I_1$$=$$I_2$$=$0, and the magnet's stored energy is $E\st{mag,0}$$=$$\frac{1}{2} L\st{m} I\st{m}^2$.
	When the ESC units are triggered, $E\st{mag,0}$ plus the energy initially stored in the ESC units is redistributed among the three coils.
	Part of the magnet's stored energy is temporarily shifted from the magnet self-inductance to the mutual-coupling energy between the magnet and the ESC coils ($M\st{m,1} I\st{m} I_1 + M\st{m,2} I\st{m} I_2$).
	
	It is important to stress that the magnetic energy is not instantaneously dissipated at this stage, but it mainly remains in the magnet volume.
	The only energy dissipation occurs in the room-temperature part of the circuit (see R$\st{circuit}$, R$\st{crowbar}$, D$\st{crowbar}$ in Figure~\ref{fig_circuit}), in the magnet coil resistance R$\st{m}$, in the ESC coil resistances R$_1$ and R$_2$, and in the ESC unit resistances (R$\st{c1}$, R$\st{c2}$) and thyristors (T$\st{c1}$, T$\st{c2}$).

	\subsection{Governing equations}
	The set of equations governing an ESC transient are presented to gain insights on the key operating parameters affecting the introduced current changes and transient losses.
	The equations presented in this section are solved numerically with the STEAM-LEDET program~\cite{LEDET_Cryogenics,thesis_CLIQ,websiteSTEAM_new} version 2.03.02 to simulate the electro-magnetic and thermal transient occurring during an ESC discharge.
	
	\subsubsection{System with one ESC circuit}
	A~magnet protected by one ESC coil is considered (see Figure~\ref{fig_circuit} if ESC circuit~2 were not present).
	Upon quench detection, the power supply's switch SW is opened and the ESC unit is triggered.
	The current through PS is nil and the diode D$\st{rev}$ is polarized in blocking direction since a negative voltage is developed across it.
	The magnet current $I\st{m}$~[A] flows through the power supply's crowbar.
	The system electrodynamics is described by the following set of equations:
	\begin{equation}
	\label{eq_01}
	\fl
	\cases{
		L\st{m} \frac{dI\st{m}}{dt} + (R\st{m} + R\st{crowbar} + R\st{circuit}) I\st{m} + U\st{crowbar} + M\st{m,1} \frac{dI\st{1}}{dt} = 0 \\
		L\st{1} \frac{dI\st{1}}{dt} + R\st{1} I\st{1} + M\st{m,1} \frac{dI\st{m}}{dt} + R\st{c1} I\st{c1} + U\st{T,c1} + U\st{c1} = 0 \\
		I\st{c1} = C\st{c1} \frac{dU\st{c1}}{dt} \\
		I\st{1} = I\st{c1} + I\st{D1} \\
	}
	\end{equation}
	where $L\st{m}$~[H] and $L\st{1}$~[H] are the self-inductances of the magnet and ESC coils, $M\st{m,1}$~[H] the mutual inductance between them, $R\st{m}$~[$\Omega$] and $R\st{1}$~[$\Omega$] their resistances, $R\st{circuit}$~[$\Omega$] and $R\st{crowbar}$~[$\Omega$] the resistances of the warm parts of the circuit and the power-supply crowbar, $U\st{crowbar}$~[V] the voltage across the crowbar's diode or thyristor, $U\st{c1}$~[V] the voltage across the ESC unit, $C\st{c1}$~[F] and $R\st{c1}$~[$\Omega$] its capacitance and internal resistance, and $U\st{T,c1}$ the voltage across the unit's triggering system.
	$L\st{m}$ and $L\st{1}$ are non-linear since they depend on the magnet current due to saturation of ferromagnetic elements surrounding the coils and on the presence of transient effects in the superconductor, such as inter-filament and inter-strand coupling currents~\cite{ASC2016_IFCC_LEDET} and filament magnetization~\cite{ASC2012_Magnetization,PersistentCurrents_LEDET_Report2020,PersistentCurrents_applications_LEDET_Report2021,10439649}.
	The non-linear part of the crowbar impedance and of the ESC unit are usually much smaller than the resistances in the circuit and is neglected, i.e. $U\st{crowbar}$$\approx$0 and $U\st{T,c1}$$\approx$0.
	
	At the moment of the ESC unit triggering ($t$=$t\st{tr}$), the current through the magnet coil is $I_0$~[A], the voltage across the ESC unit is its initial charging voltage $U\st{c1}(t$=$t\st{tr})$=$U_0$~[V], and the currents through the ESC coil and its diode are zero.
	Thus, at this moment the rates of change of the magnet and ESC coil currents are
	\begin{equation}
	\label{eq_02}
	\fl
	\cases{
		\frac{dI\st{m}}{dt}(t=t\st{tr})
		= \xi\st{m,1} \left( \frac{U_0}{L\st{1}} \frac{M\st{m,1}}{L\st{m}}
		-\frac{I_0}{\tau\st{m}} \right) \\
		\frac{dI\st{1}}{dt}(t=t\st{tr})
		= \xi\st{m,1} \left( -\frac{U_0}{L\st{1}}
		+\frac{I_0}{\tau\st{m}} \frac{M\st{m,1}}{L\st{1}} \right)
	}
	\end{equation}
	where the characteristic discharge time of the magnet circuit $\tau\st{m}$=$L\st{m}/(R\st{m}$$+$$R\st{crowbar}$$+$$R\st{circuit})$~[s] and the coupling proportionality factor $\xi\st{m,1}$=$1/(1 -k\st{m,1}^2)$ are defined.
	
	The expressions of both current changes contain a term proportional to $U_0/L\st{1}$, which represents the current change induced by ESC, and a term proportional to $I_0/\tau\st{m}$, which represents the current change induced by the circuit resistance.
	Both current changes are proportional to $\xi\st{m,1}$, which strongly depends on the magnetic coupling between the magnet and ESC coil.
	Its value is about 2.0, 2.8, and 5.3 for a $k\st{m,1}$ of 0.7, 0.8, and 0.9, respectively.
	
	The magnetic-field change resulting from the magnet and ESC coil current changes is $dB\st{a}/dt$$=$$f\st{m} dI\st{m}/dt + f\st{1} dI\st{1}/dt$, where $f\st{m}$~[T/A] and $f\st{1}$~[T/A] are the magnetic-field transfer functions of the magnet and ESC coil, respectively.
	When a magnetic-field change is applied to the superconductor, transient losses arise.
	The most relevant loss contribution depends on the superconductor type and conductor specifications.
	In low-temperature superconductors for accelerator magnets, it is expected that the dominating loss developed during an ESC transient is the inter-filament coupling loss (IFCL)~\cite{thesis_CLIQ}.
	In this contribution, other loss types such as inter-strand coupling loss and hysteresis loss are neglected.
	
	IFCL occurs when currents are driven between twisted superconducting filaments by the change of magnetic flux linked to the loops they form~\cite{Morgan70,Carr74,thesis_Arjan,Wilson,Russenschuck}.
	These inter-filament coupling currents (IFCC) generate a field $B\st{if}$~[T] that opposes to $dB\st{a}/dt$.
	The total magnetic-field change is the sum of the applied and induced fields, $dB\st{t}/dt$$=$$dB\st{a}/dt +dB\st{if}/dt$.
	
	The field induced by the IFCC can be calculated~as
	\begin{eqnarray}
	\label{eq_03}
	\fl B\st{if} = -\tau\st{if} \frac{dB\st{t}}{dt} ,
	\end{eqnarray}
	where the characteristic time constant $\tau\st{if}$~[s] is identified.
	For round composite conductors, its value is calculated as~\cite{Morgan70,Carr74,thesis_Arjan,Wilson}:
	\begin{eqnarray}
	\label{eq_04}
	\fl \tau\st{if}=\frac{\mu_0}{2}  \left(\frac{l\st{f}}{2\pi}\right)^2  \frac{1}{\rho\st{eff}} ,
	\end{eqnarray}
	with $l\st{f}$~[m] the filament twist-pitch, $\rho\st{eff}$~[$\Omega$m] the effective transverse resistivity of the strand matrix~\cite{Carr75,Zhou_2012_TransverseResistivity,fRoEff_Explanation_1,fRoEff_Explanation_2}, and $\mu_0$$=$4$\pi 10^{-7}$~Tm/A the magnetic permeability of vacuum.
	
	During transients with constant or slowly changing $dB\st{a}/dt$, the power per unit of wire volume generated by inter-filament coupling loss is~\cite{thesis_Arjan}
	\begin{equation}
	\label{eq_05}
	\fl P\st{if}'''
	=\left(\frac{l\st{f}}{2\pi}\right)^2 \frac{1}{\rho\st{eff}} \left(\frac{dB\st{t}}{dt}\right)^2 
	=\frac{2}{\mu_0} \tau\st{if} \left(\frac{dB\st{t}}{dt}\right)^2
	.
	\end{equation}
	During transients when $dB\st{a}/dt$ changes rapidly with respect to $\tau\st{if}$, a more detailed calculation of $P\st{if}'''$ is required~\cite{Turck79,Turck82,Louzguiti_2016}.
	
	\subsubsection{System with two ESC circuits}
	In order to increase the redundancy of the ESC protection system, two or more ESC circuits can be included.
	As~an example, a system composed of a magnet coupled to two ESC coils is analyzed (see Figure~\ref{fig_circuit}).
	The following set of equations, where the quantities with suffix "2" refer to the second ESC circuit, describes the electrical transient in the magnet and ESC circuits:
	\begin{equation}
	\label{eq_06}
	\fl
	\cases{
		L\st{m} \frac{dI\st{m}}{dt} + (R\st{m} + R\st{crowbar} + R\st{circuit}) I\st{m} + U\st{crowbar} + M\st{m,1} \frac{dI\st{1}}{dt}  + M\st{m,2} \frac{dI\st{2}}{dt} = 0 \\
		L\st{1} \frac{dI\st{1}}{dt} + R\st{1} I\st{1} + M\st{m,1} \frac{dI\st{m}}{dt}  + M\st{1,2} \frac{dI\st{2}}{dt} + R\st{c1} I\st{c1} + U\st{T,c1} + U\st{c1} = 0 \\
		L\st{2} \frac{dI\st{2}}{dt} + R\st{2} I\st{2} + M\st{m,2} \frac{dI\st{m}}{dt}  + M\st{1,2} \frac{dI\st{1}}{dt} + R\st{c2} I\st{c2} + U\st{T,c2} + U\st{c2} = 0 \\
		I\st{c1} = C\st{c1} \frac{dU\st{c1}}{dt} \\
		I\st{c2} = C\st{c2} \frac{dU\st{c2}}{dt} \\
		I\st{1} = I\st{c1} + I\st{D1} \\
		I\st{2} = I\st{c2} + I\st{D2} \\
	}
	\end{equation}
	
	Analytical expressions of the current changes are impractical to report for a general case.
	However, the relevant case of symmetric, identical ESC coils connected to identical protection units can provide useful insights, and is described here.
	In this case, the current changes at the moment of the ESC unit triggering are
	\begin{equation}
	\label{eq_07}
	\fl
	\cases{
		\frac{dI\st{m}}{dt}(t=t\st{tr})
		= \xi\st{m,1,2} \left[ 2 \frac{U_0}{L\st{1}} \frac{M\st{m,1}}{L\st{m}}
		-\frac{I_0}{\tau\st{m}} \left( 1 +k\st{1,2} \right) \right] \\
		\frac{dI\st{1}}{dt}(t=t\st{tr})
		= \frac{dI\st{2}}{dt}(t=t\st{tr})
		= \xi\st{m,1,2} \left( -\frac{U_0}{L\st{1}}
		+\frac{I_0}{\tau\st{m}} \frac{M\st{m,1}}{L\st{1}} \right)
	}
	\end{equation}
	where the coupling proportionality factor $\xi\st{m,1,2}$=$(1-k\st{1,2})/[1 -2 k\st{m,1}^2 (1-k\st{1,2}) -k\st{1,2}^2]$ is defined, with $k\st{1,2}$$=$$M\st{1,2}/\sqrt{L\st{1}L\st{2}}$.
	These expressions are qualitatively the same as those presented in equation~\ref{eq_02}.
	
	Not all combinations of $k\st{m,1}$ and $k\st{1,2}$ are physically possible.
	In fact, the mutual coupling factors between three coils $k_{12}$, $k_{13}$, and $k_{23}$ must satisfy the condition
	\begin{eqnarray}
	\label{eq_08}
	\fl
	k_{12}^2 + k_{13}^2 + k_{23}^2 - 2k_{12}k_{13}k_{23} \leq 1 ,
	\end{eqnarray}
	which for the symmetric 2-ESC-coil system yields
	\begin{eqnarray}
	\label{eq_09}
	\fl
	2 k\st{m,1}^2 + k\st{1,2}^2 - 2k\st{m,1}^2k\st{1,2} \leq 1 .
	\end{eqnarray}

	\section{Characteristic ESC transient}\label{section_03}
	The ESC concept is illustrated by discussing a simulated transient in a 12~T, Nb$_3$Sn racetrack dipole magnet named Short Model Coil (SMC)~\cite{Bajko:1450907,Regis:1229434}, whose parameters are summarized in Table~\ref{tab01}.
	\begin{table}[!t]
		\caption{Main Magnet and Conductor Parameters~\cite{Bajko:1450907,Regis:1229434}}
		\label{tab01}
		\centering
		\begin{tabular}{ l c c }
			\hline \hline
			Parameter & Unit & Value \\
			\hline
			Nominal current, $I\mathrm{_{nom}}$ & A & 14500 \\
			Peak field in the conductor at $I\mathrm{_{nom}}$ & T & 12.5 \\
			Operating temperature & K & 1.9 \\
			Differential inductance per unit length at $I\mathrm{_{nom}}$ & mH/m & 3.9 \\
			Number of turns & - & 2x35 \\
			Insulated coil cross-section & mm$^2$ & 3404 \\ 
			\hline
			Bare cable width & mm & 14.847 \\
			Bare cable height & mm & ~1.305 \\
			Number of strands & - & 40 \\
			Strand diameter & mm & 0.7 \\
			Cu/no-Cu ratio & - & 1.106 \\
			Insulation thickness & $\mu$m & ~150 \\
			\hline
			\hline
		\end{tabular}
	\end{table}
	The magnet reaches a bore magnetic field of 12~T when operated at a nominal current of 14.5~kA and at a temperature of 1.9~K.
	
\begin{figure}[!t]
		\centering
		\includegraphics[trim={0 0 0 0cm},clip,width=0.8\textwidth]{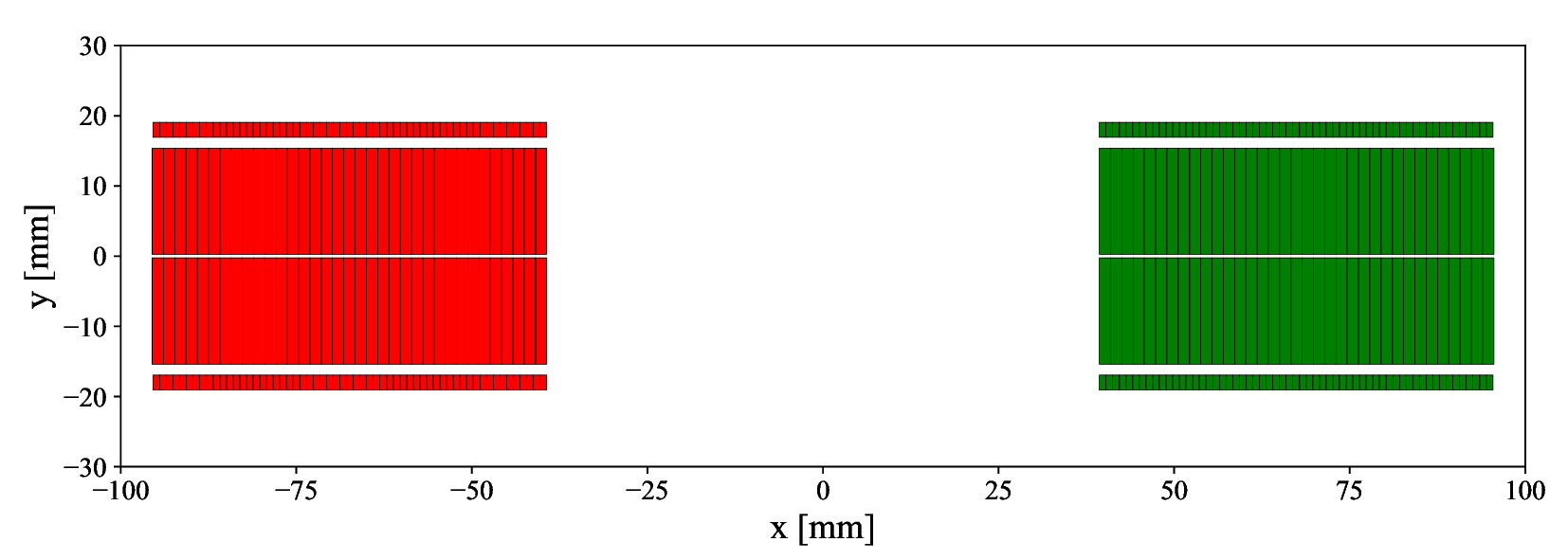}
		\caption{Cross-section of the SMC magnet coils and two ESC coils made of a 2x0.8~mm$^2$ rectangular wire.
			The colors indicate the current polarities corresponding to the positive currents $I\st{m}$, $I_1$, and $I_2$ in Figure~\ref{fig_circuit}.
			\label{fig01}
		}
	\end{figure}

	The windings cross-section including the magnet coil and two ESC coils is shown in Figure~\ref{fig01}.
	%
	Each ESC coil has 61 turns and is made of 2.0x0.8~mm$^2$ bare rectangular copper wire with 75~$\mu$m enamel insulation, as indicated in Table~\ref{tab02}.
	\begin{table}[!t]
		\caption{Main ESC Coil and Conductor Parameters}
		\label{tab02}
		\centering
		\begin{tabular}{ l c c }
			\hline \hline
			Parameter & Unit & Value \\
			\hline
			Self-inductance $L_1$ per unit length at $I\mathrm{_{nom}}$ & mH/m & 3.3 \\
			Mutual coupling factor between $L_1$ and $L\st{m}$ & - & 0.79 \\
			Mutual coupling factor between $L_1$ and $L_2$ & - & 0.51 \\
			Number of turns & - & 61 \\
			Insulated coil cross-section & mm$^2$ & 475 \\
			\hline
			Bare wire dimensions & mm$^2$ & 2.0x0.8 \\
			Insulation thickness & $\mu$m & ~75 \\
			\hline
			\hline
		\end{tabular}
	\end{table}
	The two coils are placed on the top and bottom of the magnet coils and their turns cover the entire magnet winding pack, but not the pole piece around which the magnet coil is wound.
	Each ESC coil occupies a cross-section that is about 7\% of the magnet coil cross-section.
	Additional insulation layers are included between the magnet cable insulation and the ESC wire insulation.
	The distance between the bare conductors of the magnet and ESC coil is about 1.8~mm. 
	The mutual coupling parameters are $k\st{m,1}$$\approx$0.79 and $k\st{1,2}$$\approx$0.51, which result in a proportionality factor $\xi\st{m,1,2}$$\approx$3.6.
	
	\subsection{Simulated ESC transient}\label{transient_ESC}
\begin{figure}[!t]
		\centering
		\includegraphics[trim={0 0 0 0.9cm},clip,width=0.6\textwidth]{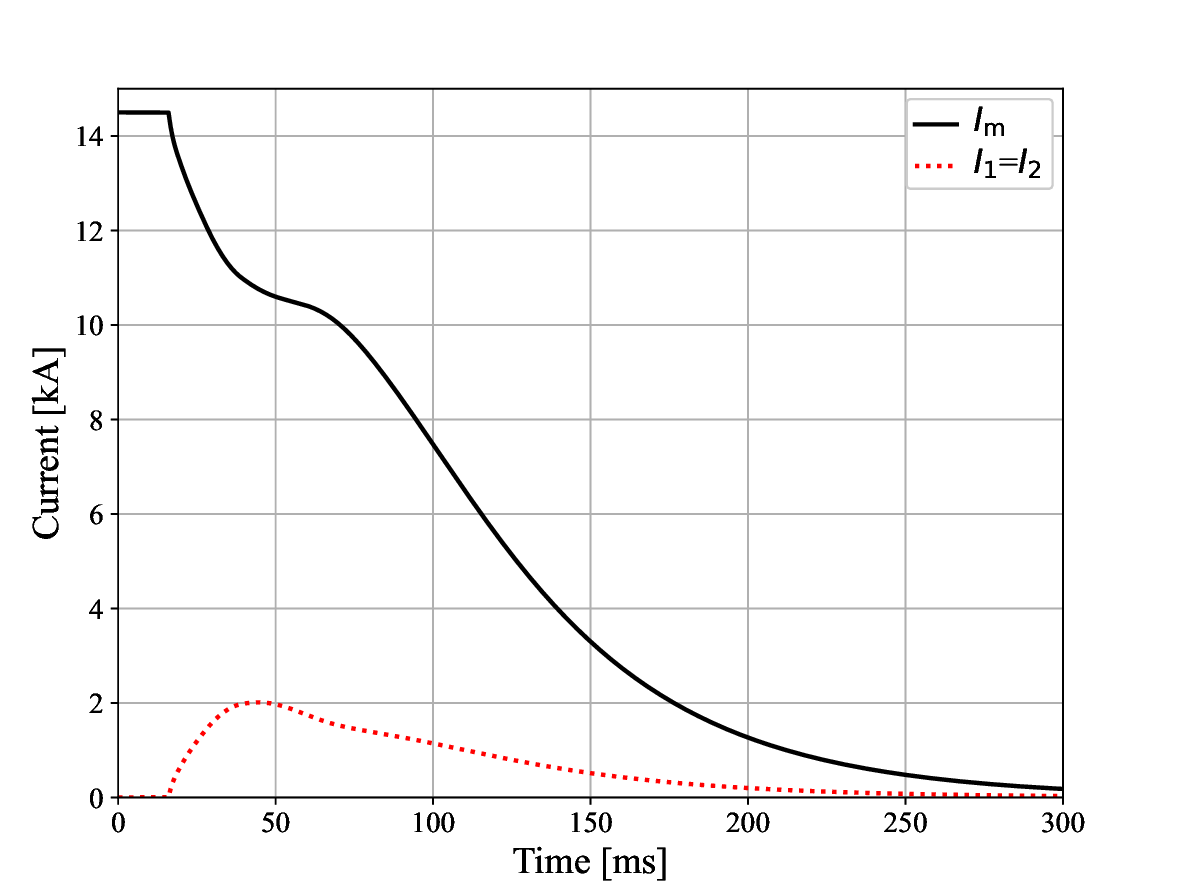}
		\caption{Simulation of the transient following the activation of an ESC system on 15~m long SMC magnet.
			Current flowing in the magnet coil and the ESC coils, versus time.
			\label{fig02}
		}
	\end{figure}

	In order to demonstrate the ESC quench protection performance on a full-scale magnet the case of a 15~m long SMC magnet is considered, even though its actual magnetic length is 0.3~m and there is no plan to build a longer version of the magnet.
	The electro-magnetic and thermal transients developed during an ESC discharge following a quench occurring at $t$=0 at the nominal current of 14.5~kA are simulated with the STEAM-LEDET software, which includes ohmic loss, thermal diffusion, and inter-filament coupling loss~\cite{LEDET_Cryogenics,thesis_CLIQ,websiteSTEAM_new}.
	The simulated currents flowing in the magnet and ESC coils after triggering two 30~mF, 1~kV ESC units at $t$=16~ms are shown in Figure~\ref{fig02}.
	Due to symmetry the currents through the two ESC coils are identical.
	
	Upon ESC unit triggering, a positive current change is imposed in the ESC coils, whose peak currents reach about 2~kA.
	Simultaneously, a negative current change is imposed in the magnet coils, which reaches a few hundreds kA/s. 
	During this phase, most magnet turns are still superconducting and inter-filament coupling currents develop, which result in a very significant reduction of the magnet differential inductance~\cite{ASC2016_IFCC_LEDET}.
	
	The fast current changes cause large $dB\st{t}/dt$ in the magnet windings.
	Thus, very high inter-filament coupling loss is generated in the superconductor, which quickly heats up.
	As~a result, the entire magnet coil volume is transitioned to the normal state between 2 and 9~ms after ESC activation.
	The rapid increase of the magnet coil resistance helps maintaining a negative $dI\st{m}/dt$ and a positive $dI_1/dt$ even as the voltage across the ESC unit decreases.
	
	About 30~ms after the ESC triggering, the resistive voltages $R_1 I_1$ and $R_2 I_2$ built up in the ESC coils are high enough to induce negative $dI_1/dt$ and $dI_2/dt$.
	Consequently, due to magnetic coupling the amplitude of $dI\st{m}/dt$ decreases as part of the energy is shifted back to the magnet coils (see 35~ms$\leq$$t$$\leq$60~ms in Figure~\ref{fig02}).
	The diodes across the ESC units prevent them from significantly charging with opposite polarity with respect to their initial charging voltage.
	In the last part of the transient ($t$$\geq$80~ms) both the magnet and ESC coil currents are discharged semi-exponentially with a similar discharge time.
	
\begin{figure}[!t]
		\centering
		\includegraphics[trim={0 0 0 0.9cm},clip,width=0.6\textwidth]{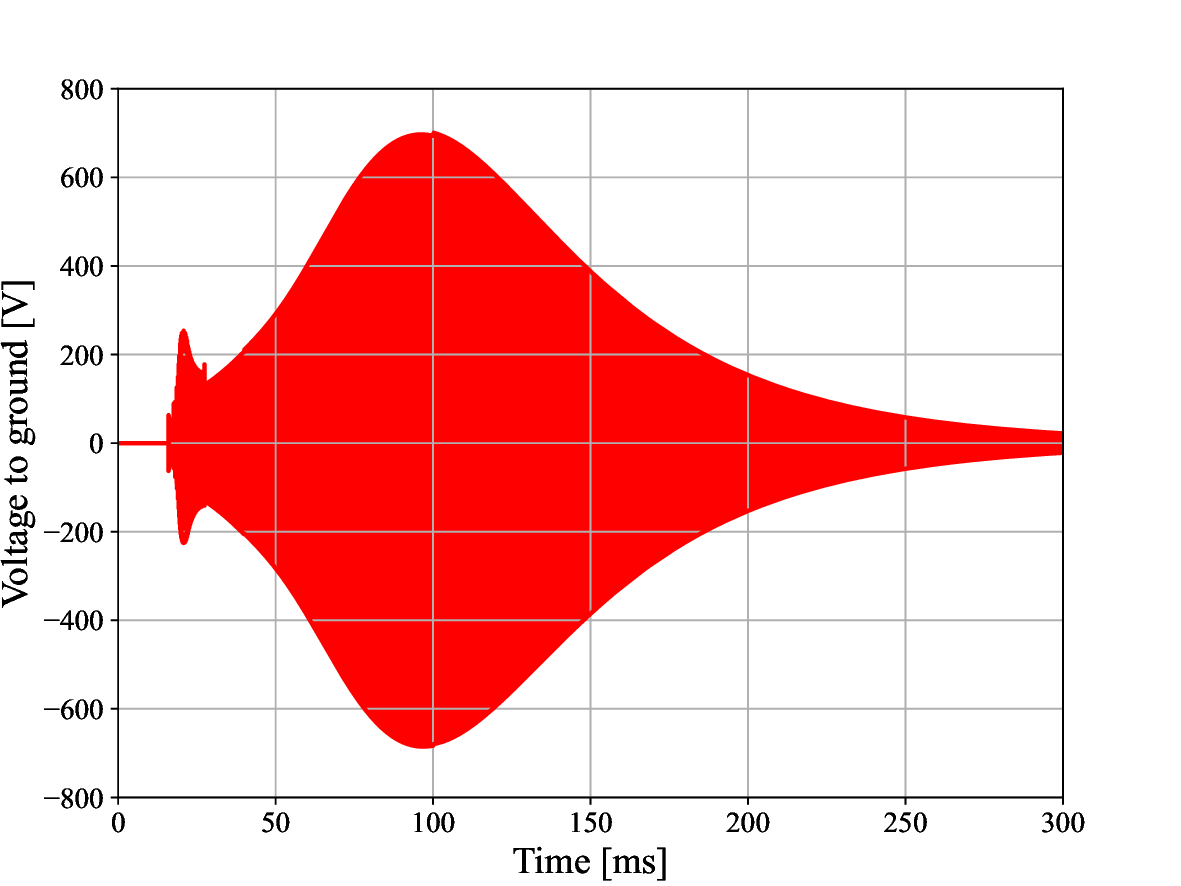}
		\caption{Simulation of the transient following the activation of an ESC system on a 15~m long SMC magnet.
			Voltages to ground in turns of the magnet coil, versus time.
			\label{fig03}
		}
	\end{figure}

	The simulated voltages to ground developed in the magnet windings are shown in Figure~\ref{fig03}.
	At the moment of ESC activation, while the voltage to ground in the ESC coil windings reaches $\pm$500~V, i.e. half of the units' charging voltages, the voltage to ground in the magnet coil windings is lower than 65~V.
	A~few milliseconds after, when the coil turns start quenching, the voltage to ground increases due to uneven resistive and inductive internal voltage distribution, and it reaches its maximum value 80~ms after ($t$$\approx$96~ms).
	The peak value of about 700~V is relatively low for a 15~m long magnet with peak magnetic field of 12~T.
	This result can only be achieved due to the very homogeneous quench in the windings.
	
\begin{figure}[!t]
		\centering
		\includegraphics[trim={0 0 0 0.9cm},clip,width=0.6\textwidth]{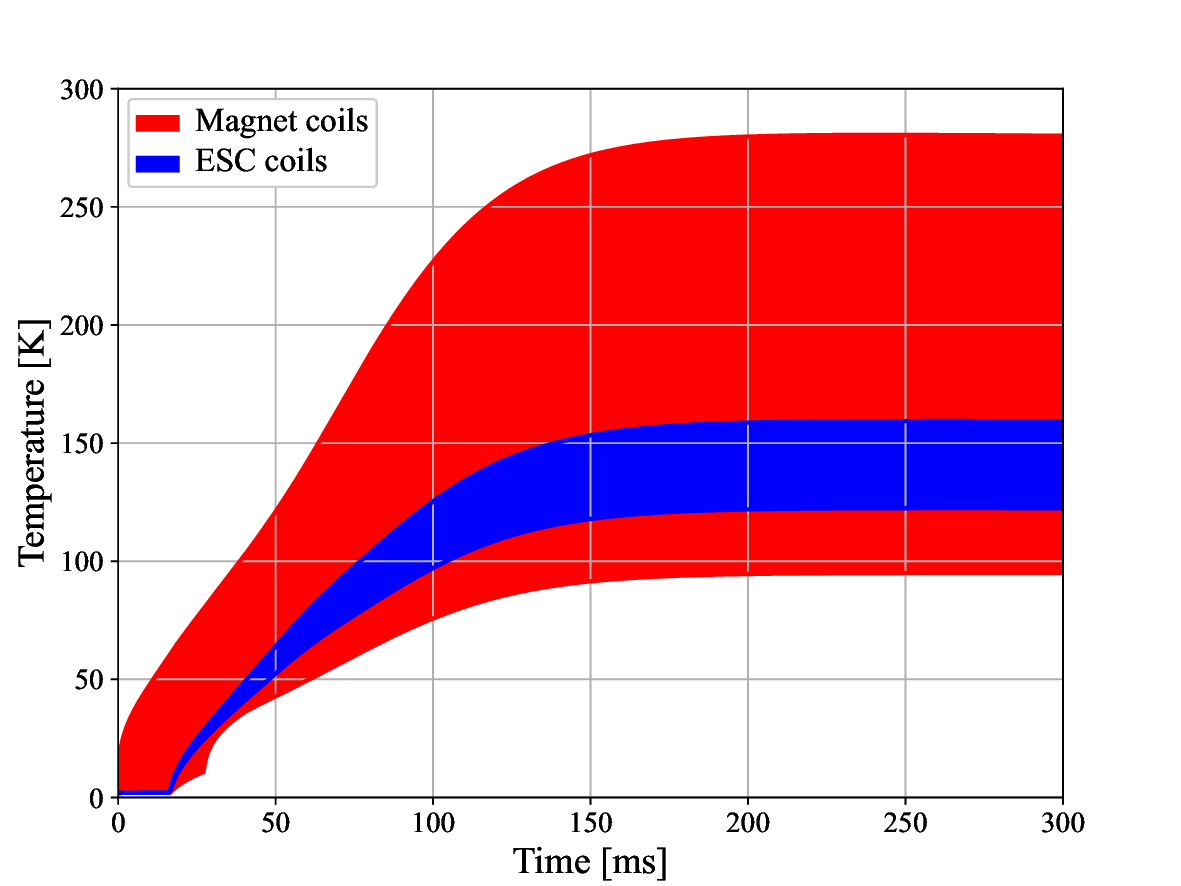}
		\caption{Simulation of the transient following the activation of an ESC system on a 15~m long SMC magnet.
			Temperature in the windings of the magnet and ESC coils, versus time.
			\label{fig03b}
		}
	\end{figure}

	The temperature $T$~[K] evolution in the magnet and ESC coils is shown in Figure~\ref{fig03b}.
	The hot-spot starts heating up at $t$=0 when the quench starts, and reaches $T$$\approx$280~K.
	When calculated under adiabatic assumptions, i.e. neglecting the heat diffusion to adjacent turns, the calculated hot-spot temperature $T\st{hot}$~[K] is 290~K.
	
	The temperature in the ESC windings reaches 160~K.
	Their heating is dominated by the energy transferred from the magnet to the ESC coils, which is about 19\% of the magnet's stored energy~$E\st{mag,0}$.
	The energy stored in the ESC units is less than 0.5\% of $E\st{mag,0}$, which is only enough to heat up the ESC conductor to less than 40~K.
	Thermal diffusion between magnet and ESC coils plays no role since they are separated by almost 2~mm of insulating material, and the quench transient lasts only about 300~ms.
	
\begin{figure}[!t]
		\centering
		\includegraphics[trim={0 0 0 0cm},clip,width=0.8\textwidth]{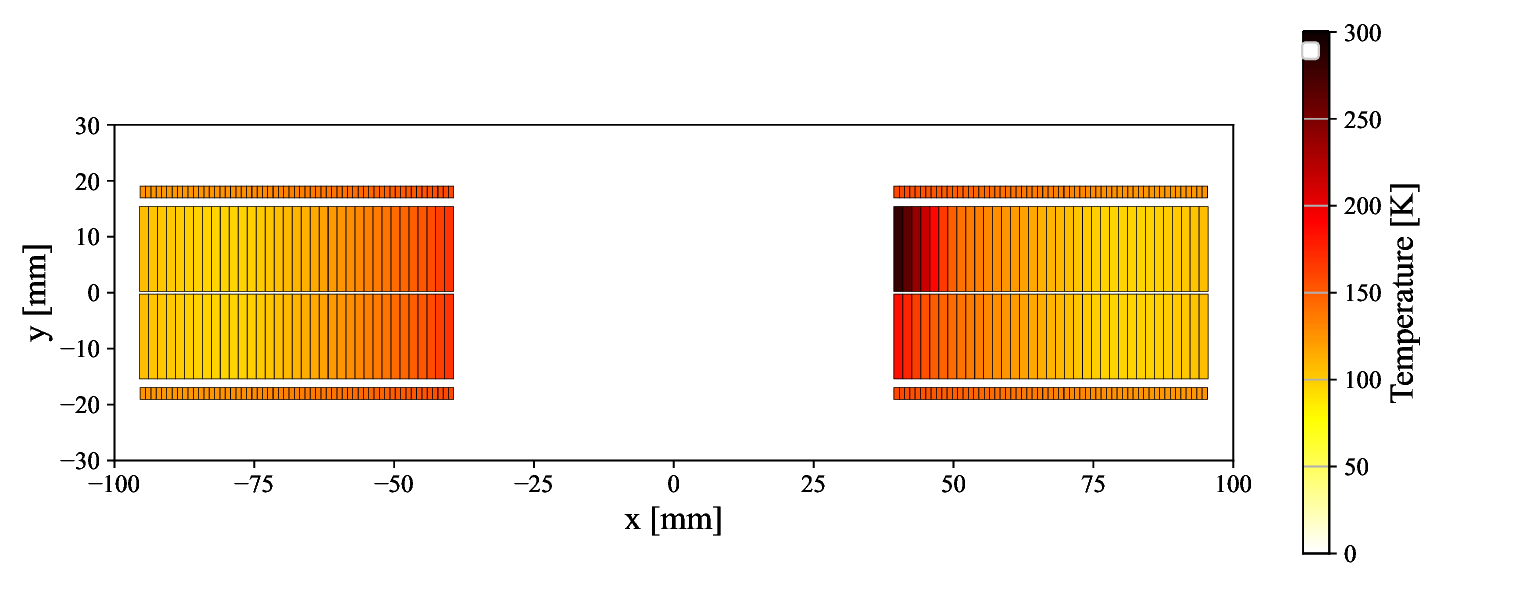}
		\caption{Temperature distribution in the windings of the magnet and ESC coils at the end of the ESC discharge ($t$=300~ms).
			\label{fig04}
		}
	\end{figure}

	The temperature distribution in the coil cross-section at the end of the transient ($t$=300~ms) is shown in Figure~\ref{fig04}.
	Most of the magnet and ESC turns reach a temperature between 95 and 170~K.
	The $T$ profile is remarkably uniform considering that the magnetic field spans from 0~T to more than 12~T in the conductor, and hence the magneto-resistance and the energy margin to quench are significantly inhomogeneous.
	
	\subsection{Synergy between ESC voltage and coil resistive voltage}
\begin{figure}[!t]
		\centering
		\includegraphics[trim={0 0 0 0cm},clip,width=0.6\textwidth]{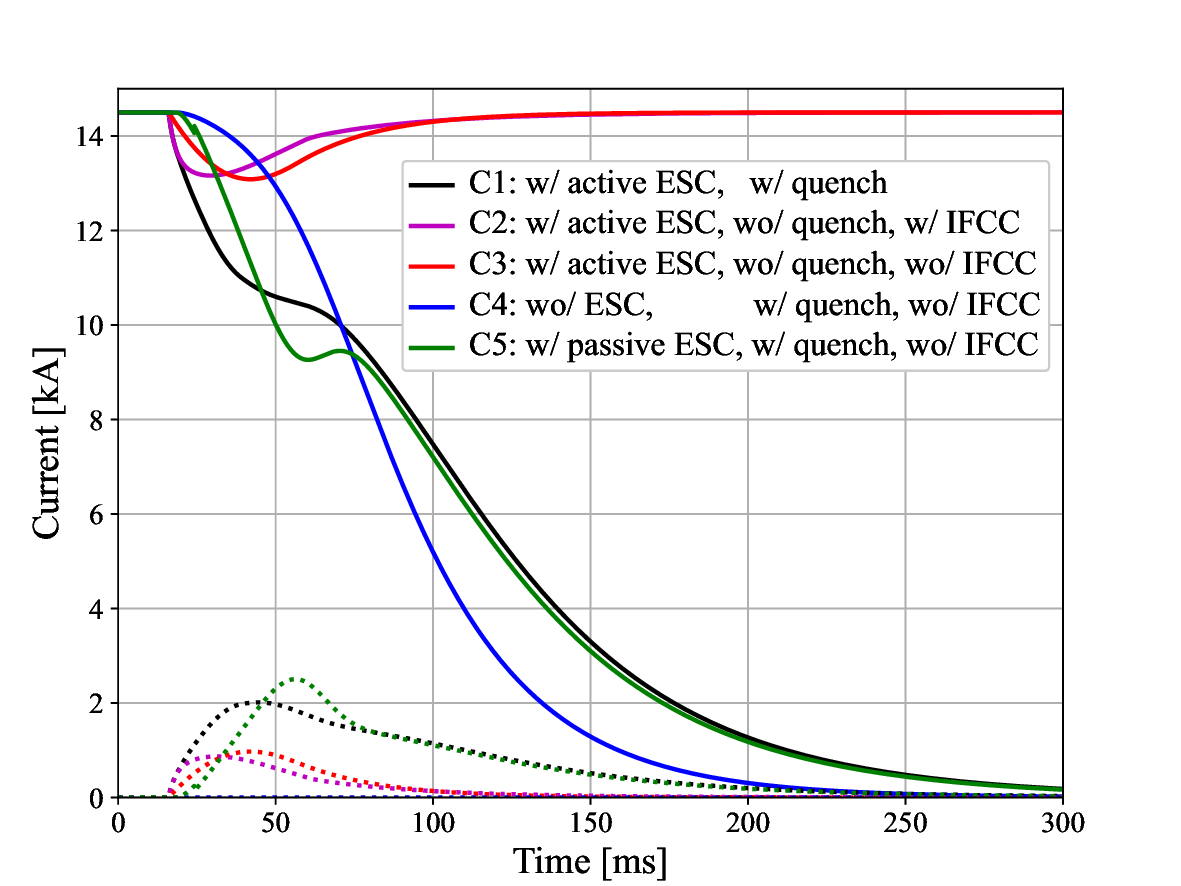}
		\caption{Simulated magnet and ESC coil currents, versus time, for the five cases defined in Table~\ref{tab03}.
			\label{fig05}
		}
	\end{figure}

	To better illustrate the synergy in action between the voltages imposed by the ESC units and by the built-up of magnet coil resistance, five simulations are performed including individually different physical features.
	The assumptions of each simulation case are summarized in Table~\ref{tab03}, together with the main quench protection results, while the simulated currents are shown in Figure~\ref{fig05}.
	\begin{table}[!t]
		\caption{Assumptions and Results for Various Simulation Cases}
		\label{tab03}
		\centering
		\begin{tabular}{ l c c c | c c c c c}
			\hline \hline
			Case & ESC & IFCC & Quench & $f\st{ex}$~[-] & $t\st{10\%}$~[ms] & $t\st{90\%}$~[ms] & $T\st{hot}$~[K] & $U\st{g}$~[V] \\
			\hline
			C1 & active  & yes & yes & 19\% & 2 & 9 & 290 & ~701 \\
			C2 & active  & yes & no & ~0\% & - & - & $\gg$500 & ~~63 \\
			C3 & active  & no & no & ~0\% & - & - & $\gg$500 & ~~13 \\
			C4 & no      & no & yes & ~0\% & 2 & 9 & 316 & 1013 \\
			C5 & passive & no & yes & 19\% & 2 & 9 & 294 & ~742 \\
			\hline
			\hline
		\end{tabular}
	\end{table}
	Case~1 (C1) is identical to the simulation presented in Section~\ref{transient_ESC}.
	It includes ESC activation, inter-filament coupling currents, and quench state.
	
	In Case~2, the quench is disabled, and hence no resistance is built-up in the magnet coil.
	When ESC is activated, the same initial $dI\st{m}/dt$ as Case~1 is induced.
	However, since $R\st{m}$$=$0 the magnet current returns to its initial value after about 100~ms.
	The energy shifted from the magnet coils to the ESC coils is fully returned to the magnet, and hence the fraction of extracted energy $f\st{ex}$ is nil.
	The peak voltage to ground $U\st{g}$ is lower than 65~V, which is the same value as Case~1 before the magnet turns are transitioned to the normal state.
	If IFCC are also disabled (see Case~3), the peak $dI\st{m}/dt$ is reduced by about five times.
	This result shows that IFCC have a significant effect on the magnet's differential inductance in such a fast transient, and this effect must be included in any ESC simulation.
	
	If ESC is not triggered, but each magnet turn is set to quench at exactly the same time as in Case~1, no energy is extracted and the magnet is protected less effectively, i.e. the peak temperature and voltage to ground reached during the transient increase (see Case~4).
	Finally, if the magnet turns are set to quench at the same times as in Case~1 and the ESC coils are present, but not powered, the resulting extracted energy fraction, hot-spot temperature, and peak voltage to ground are very similar to Case~1 (see Case~5).
	On the contrary, the peak $dI\st{m}/dt$ is significantly reduced with respect to Case~1, where the effects of ESC and coil resistance act synergistically.
	
	\subsection{Redundancy}\label{sec_redundancy}
\begin{figure}[!t]
		\centering
		\includegraphics[trim={0 0 0 0cm},clip,width=0.6\textwidth]{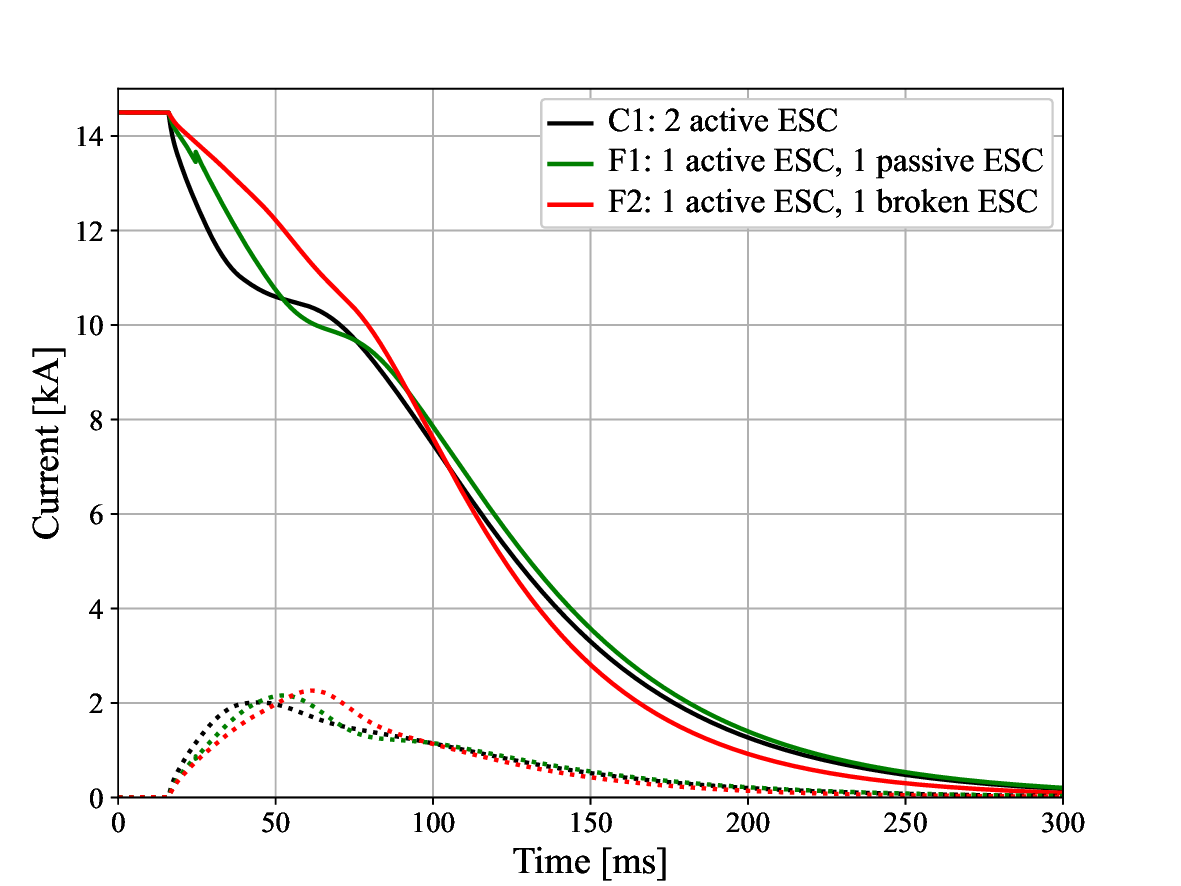}
		\caption{Effect of two failure cases on the ESC performance.
			Simulated magnet and ESC coil currents, versus time, for the reference Case~1 in Table~\ref{tab02} and the two failure cases defined in Table~\ref{tab04}.
			\label{fig07}
		}
	\end{figure}

	One advantage of the ESC system is the fact that it is straightforward to add redundancy to the system.
	In fact, multiple ESC circuits can be integrated in the magnet design, and the system can be designed to protect the magnet in the case one or more units fail.
	The simulated current in the case of failure in open circuit of one out of two ESC units is shown in Figure~\ref{fig07}, and compared to the reference Case~1.
	The $dI\st{m}/dt$ just after triggering ESC is about half of that reached in Case~1.
	Since the two ESC coils are mutually coupled, a current is induced also through the coil connected to the failing ESC unit and through the diode installed across it.
	As~a result, an acceptable quench protection performance is achieved, as shown in Table~\ref{tab04}, where the simulation results are summarized.
	\begin{table}[!t]
		\caption{Impact of Failure Cases on Quench Protection Performance}
		\label{tab04}
		\centering
		\begin{tabular}{ l c c | c c c c c}
			\hline \hline
			Case & ESC 1 & ESC 2 & $f\st{ex}$~[-] & $t\st{10\%}$~[ms] & $t\st{90\%}$~[ms] & $T\st{hot}$~[K] & $U\st{g}$~[V] \\
			\hline
			C1 & active  & active  & 19\% & 2 & ~9 & 290 & ~701 \\
			F1 & active  & passive & 18\% & 5 & 21 & 325 & 1219 \\
			F2 & active  & broken  & 10\% & 5 & 41 & 364 & 1954 \\
			\hline
			\hline
		\end{tabular}
	\end{table}
	90\% of the winding pack is transitioned to the normal state in 21~ms, and almost the same fraction of $E\st{mag,0}$ as in Case~1 can be extracted from the magnet.
	The hot-spot temperature reaches 325~K, i.e. 35~K higher than the reference case, and the peak voltage to ground reaches about 1200~V, i.e. almost twice higher than the reference case.
	
	In the case one out of two ESC coils is broken, or one unit fails and the diode is not present, there is no current induced in the failing ESC circuit.
	In such a case, the performance is considerably worsened.
	The hot-spot temperature and peak voltage to ground are above 350~K and almost 2~kV, respectively.
	This simplified failure analysis suggests that it would be prudent to install four ESC circuits and/or make the ESC diodes redundant to reduce the consequences of a failure.
	A~more detailed analysis is outside the scope of this work.
	
	
	\subsection{Importance of the diodes in the ESC circuits}\label{sec_D_ESC}
	The advantage of including the diodes D$_1$ and D$_2$ in the ESC circuits (see Figure~\ref{fig_circuit}) is fourfold.
	First, it prevents developing high voltage across the ESC unit, and hence across the coil in the cases where the voltage induced by the magnet coil resistance build-up is much higher than the ESC-induced voltage.
	Second, it increases the system reliability by allowing a current to be driven passively in the ESC coils in the case one ESC unit fails to trigger, as shown in Section~\ref{sec_redundancy}.
	Third, it prevents the recharging of the capacitor banks C$\st{c1}$ and C$\st{c2}$ during the quench transient, which would transfer back to the magnet coils a larger part of the energy shifted to the auxiliary coils.
	Finally, it provides a path for $I_1$ and $I_2$ to flow in the cryogenic part of the circuit without flowing through the ESC current leads, which allows reducing their conductor cross-section and consequent cryogenic losses.
	
\begin{figure}[!t]
		\centering
		\includegraphics[trim={0 0 0 0cm},clip,width=0.6\textwidth]{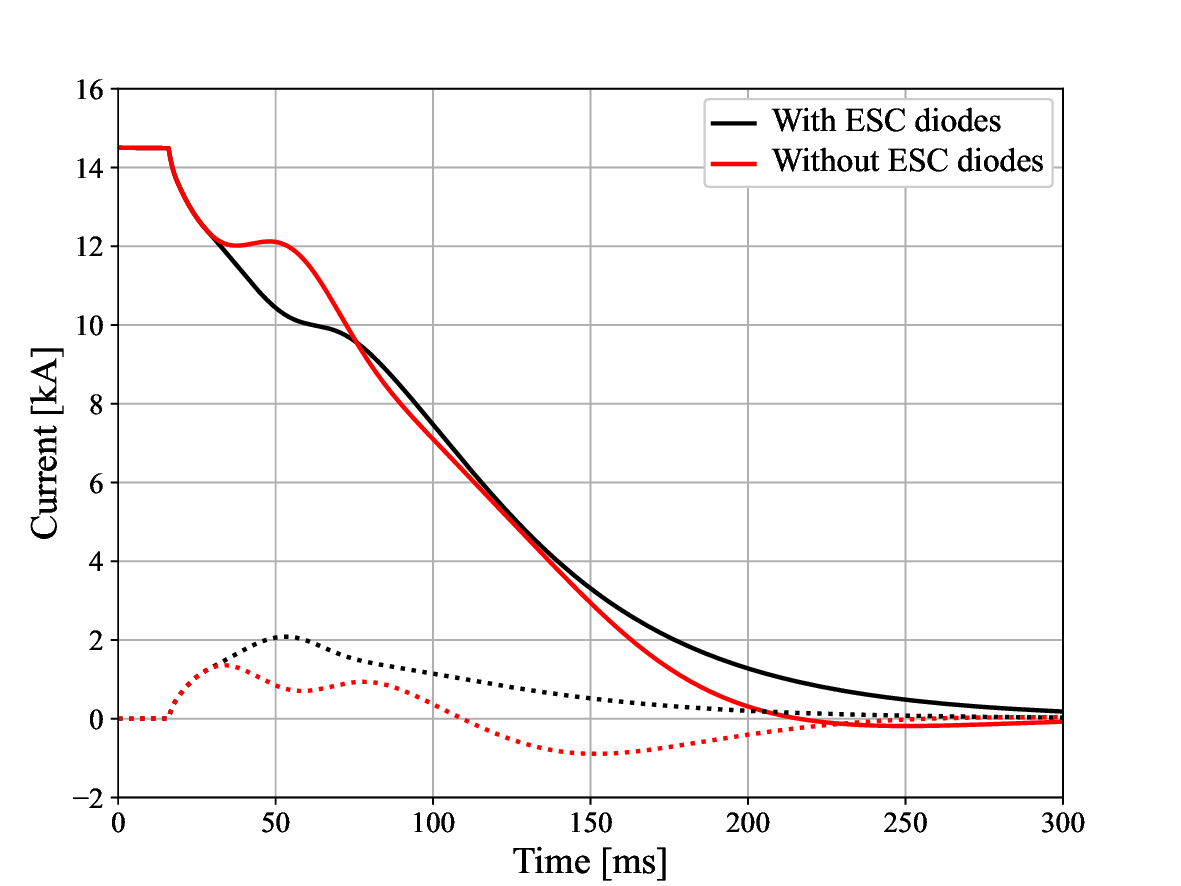}
		\caption{Simulated magnet and ESC coil currents, versus time, after triggering two 10~mF, 1~kV ESC units.
			\label{fig06}
		}
	\end{figure}

	To better illustrate the importance of including these diodes, additional simulations are run for a case where two 10~mF, 1~kV ESC units are triggered with and without the diodes installed across the units.
	The simulated currents in these two cases are shown in Figure~\ref{fig06}.
	In these simulations, the ESC units are fully discharged about 13~ms after triggering them.
	At this time, if diodes are present across the units the currents $I_1$ and $I_2$ flow through them and not through the ESC units.
	Thus, the unit components can be rated to carry a smaller thermal load.
	Furthermore, if the diodes are included in the cryogenic part of the circuit (see Figure~\ref{fig_circuit}), also the current leads connecting the ESC units to the ESC coils can be designed to carry a lower thermal load and peak current, hence reducing the thermal leak.
	
	On the contrary, if diodes are not installed across the units, they are allowed to charge with a voltage of opposite polarity with respect to their initial charging voltage, which reduces the fraction of magnet energy that is transferred to the ESC coils.
	In this example, only 4\% of $E\st{m}$ is extracted rather than 19\%.
	Furthermore, if the diodes are not installed the peak voltage to ground in the ESC windings increases from about 560~V to more than 3200~V.
	
	\subsection{Effect of initial magnet current}
\begin{figure}[!t]
		\centering
		\includegraphics[trim={0 0 0 0cm},clip,width=0.8\textwidth]{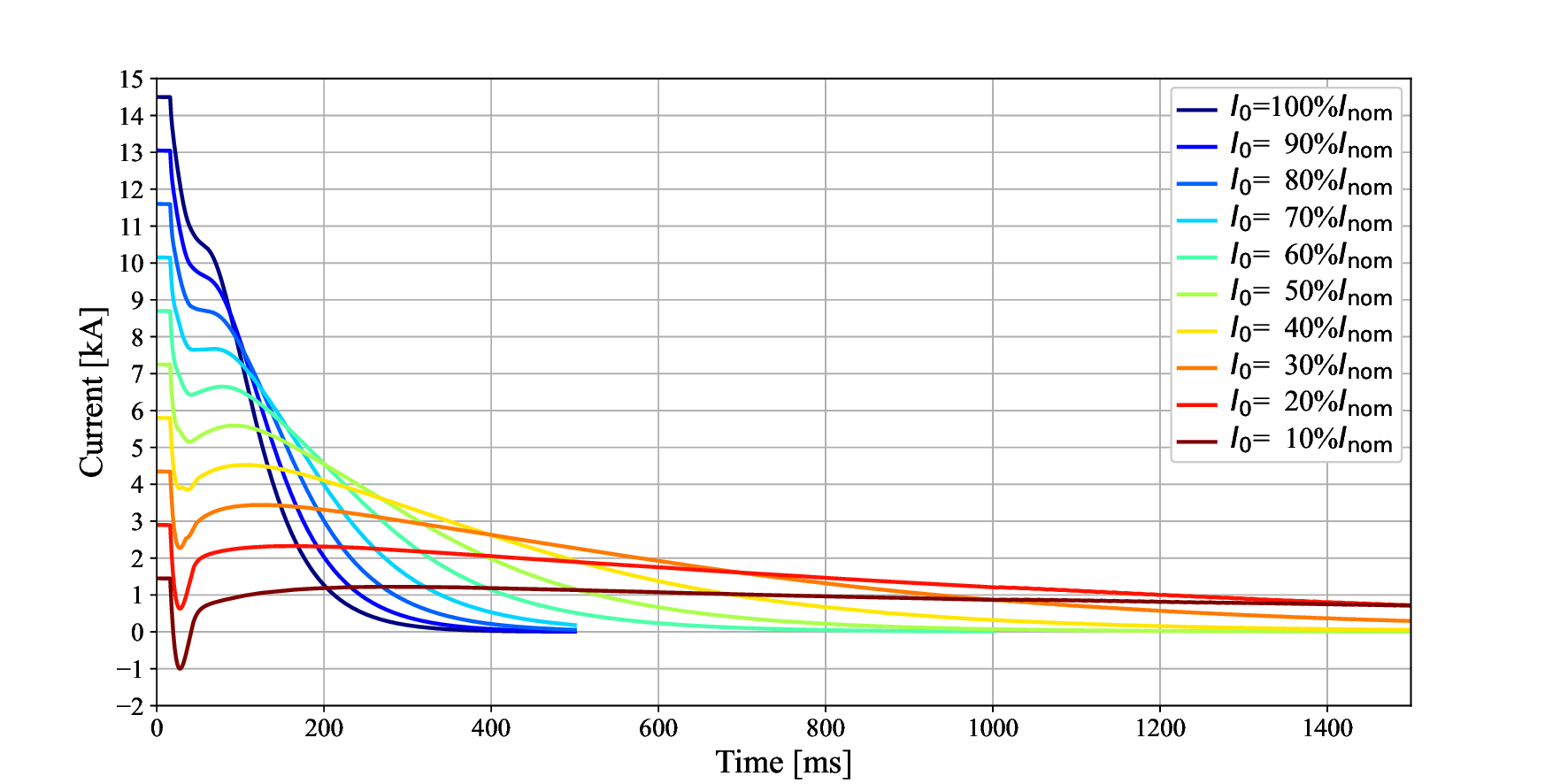}
		\caption{
			ESC performance at different initial current levels in the range of 10\% to 100\% of the nominal current.
			Simulated magnet current, versus time, after triggering two 30~mF, 1~kV ESC.
			\label{fig09}
		}
	\end{figure}

	ESC can be designed to protect the magnet at any operating current level.
	The simulated magnet currents after a quench at different initial current levels in the range of 10\% to 100\% of $I\st{nom}$ are shown in Figure~\ref{fig09}.
	The magnet conductor is transitioned to the normal state at all current levels, and the highest hot-spot temperature and peak voltage to ground is reached after the quench at the highest current.
	The lower the current level, the slower the magnet is discharged due to the slower coil resistance development, which is a consequence of the slower quench initiation and lower magneto-resistance.
	
	The current drop occurring at the ESC triggering is mostly independent of the initial magnet current.
	Thus, when $I_0$ is very low it is possible that ESC pushes the magnet transport current to negative values (see 10\%$I\st{nom}$ curve).
	This is not an issue since the diode R$\st{rev}$ assures a path for the magnet current even when $I\st{m}$$<$0.

	%
	%
	%

	
	
	\subsection{Comparison to other protection methods}\label{sec_Cu_NoCu}
\begin{figure}[!t]
		\centering
		\includegraphics[trim={0 0 0 0cm},clip,width=0.8\textwidth]{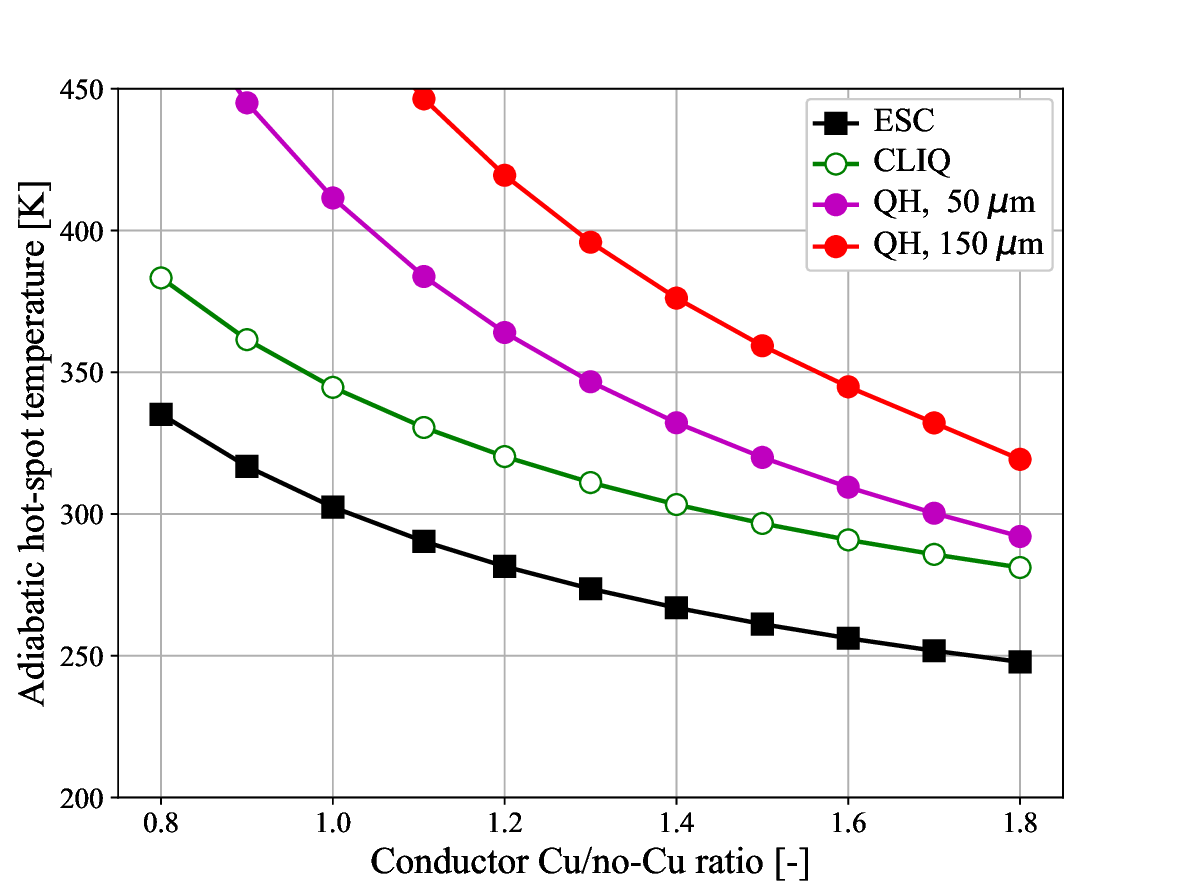}
		\caption{
			Comparison between the quench protection performance of four quench protection systems.
			Adiabatic hot-spot temperature after a quench in a 15~m long magnet at nominal current, versus conductor Cu/no-Cu ratio~$R$.
			\label{fig10}
		}
	\end{figure}

	The ESC quench protection performance is compared to that of two conventional protection methods, namely quench heaters and CLIQ.
	A~set of quench simulations is performed assuming the same magnet geometry, magnetic length, initial operating conditions, and protection unit charging voltage as the transients analyzed in the previous sections.
	The Cu/no-Cu ratio in the magnet conductor~$R$ is varied around the SMC design value of 1.106 to emphasize the potential reduction of conductor stabilizer that can be achieved when introducing ESC.
	The calculated adiabatic hot-spot temperature is plotted in Figure~\ref{fig10} as a function of $R$, for each considered quench protection method.

	Two quench heater configurations are assessed, whose strips reach a peak power density of 298~W/cm$^2$, cover more than 90\% of the magnet turns, and have heating stations covering 12.5\% of the coil length.
	The two systems have different insulation schemes: one with only a 50~$\mu$m kapton layer between QH strip and magnet conductor insulation, and one with an extra 100~$\mu$m fiber glass (G10) layer.
	Even adopting the thin-insulation design, QH can maintain the hot-spot temperature below 350~K, which is considered a safe limit with respect to permanent degradation~\cite{Ambrosio_LimitsHotSpotT}, only by increasing $R$ above the design value.
	
	A~CLIQ system including a 60~mF, 1 kV unit stores the same energy as the above-mentioned ESC system.
	It can protect this magnet, although with little margin with respect to 350~K when the reference~$R$ is adopted.
	
	ESC can maintain the hot-spot temperature below 350~K even when reducing $R$ to 0.8, which corresponds to a reduction of Cu in the conductor of about 15\% with respect to the original design, and results in a current density in the Cu cross-section higher than 2.1~kA/mm$^2$.
	This remarkable protection performance is achieved thanks to ESC combined ability of rapidly transitioning the magnet conductor to the normal state and of extracting part of the magnet's stored energy.

	\section{Discussion and Outlook}
	The proposed ESC method offers various advantages with respect to existing quench protection systems.
	It can transition the magnet conductor to the normal state very quickly since it relies on a very effective heat deposition mechanism, and it does not rely on electrical contact between the protection elements and the magnet coils, like CLIQ does, nor on physical contact, like quench heaters do.
	Thus, it can achieve very good quench protection performance with limited risk of compromising the electrical integrity of the magnet.
	
	This technology can be easily applied to magnets of different geometries, such as cos-$\theta$, block-coil, and common-coil multipole magnets, pancake coils, and solenoids.
	In order to achieve good performance, the ESC coils should have the same geometry type as the magnet coils to protect, and ideally closely follow the magnet winding.
	Top performance in terms of normal-zone generation speed can be achieved by placing ESC coils in proximity to all magnet coils.
	
	It is relatively easy to add redundancy to an ESC system.
	Multiple ESC circuits can be integrated in the magnet design by including more electrically-separated coils.
	
	Integrating ESC in the magnet design also brings disadvantages.
	The ESC copper coils occupy space in the magnet cross-section, which could otherwise be used to place superconductor.
	This drawback, which reduces the magnetic efficiency, can be offset by the achievable reduction of the content of stabilizer in the superconductor, which is shown in Section~\ref{sec_Cu_NoCu}.
	
	The interaction between ESC and other magnet sub-systems must be carefully considered.
	For example, the very high magnetic-field changes imposed by ESC cause eddy currents in all electrically-conducting elements present in the magnet volume.
	This imposes additional constraints on the design of such elements to ensure robustness.
	
	Furthermore, the electromagnetic forces developed in the magnet and ESC coils during the transient, as well as the resulting stresses, must be analyzed.
	By principle the forces acting on the magnet conductor always remain lower than their value during magnet operation, since both ESC-induced voltage and magnet-coil resistive voltage impose a negative magnet-current rate of change.
	Depending on the applications, ESC coils may need a dedicated support structure to maintain them in place.
	
	The application of ESC can enable new design spaces for future high magnetic-field magnets.
	Thanks to its ability of extracting energy from the magnet coils ESC can significantly reduce the requirements on the quench protection, which  would otherwise require to substantially increase the magnet conductor size and/or its fraction of stabilizer.
	In turn, this allows achieving a given target field and a given margin with a more compact magnet coil.
	
	Conventionally, copper or silver needs to be added to the cross-section of practical superconductors to ensure adequate superconductor stability, mechanical strength, and quench protection performance~\cite{Wilson}.
	With the introduction of ESC, the later requirement is significantly relaxed.
	In many relevant cases, decreasing the Cu/no-Cu ratio arbitrarily is not feasible due to the above-mentioned limitations, as well as the peak stresses reached in the conductor at elevated current densities.
	However, future R\&D efforts could focus on developing stable and robust strands with a Cu/no-Cu ratio lower than 0.8, or reinforced strands including an outer sheet of high-strength material~\cite{828435,6971157}.
	Alternatively, the reduction of conductor size could provide additional space in the magnet cross-section for incorporating stress-management structural elements.
	
	The ESC protection technique appears promising also for protecting high-temperature superconductor (HTS) magnets.
	The transient losses in Rutherford cables made of Bi-2212 strands~\cite{Larbalestier2014}, as well as in REBCO tapes~\cite{Selvamanickam_HTS_2012} and cables, are substantial and could be utilized as an effective heating mechanism.
	In the case of REBCO tapes, no inter-filament coupling loss occurs since they do not include superconducting filaments, and it is likely that the dominant transient loss during an ESC transient is hysteresis loss due to screening currents.
	When designing an ESC system for a REBCO magnet, it is prudent to include a detailed analysis of the screening currents developed during the fast transients~\cite{6648727,Berrospe-Juarez_2019,Shen_2020,Vargas-Llanos_2022,Bortot_2020}, which are likely very significant and could cause permanent degradation if not properly handled.
	The application of ESC to HTS magnets will be explored in future work.

	\section{Conclusion}
	A~novel quench protection method named Energy Shift with Coupling (ESC) is presented, which offers multiple advantages in terms of effectiveness, robustness, and reliability.
	The ESC system includes a set of normal conducting coils that are strongly magnetically coupled to the magnet coils, but galvanically insulated from them.
	Upon quench detection, a charged capacitor bank is discharged across each ESC coil causing a very rapid shift of magnetic energy between the magnet and ESC coils.
	This results in three beneficial effects: a sudden drop in the magnet current, which immediately reduces ohmic loss in the magnet coil hot-spot, a high magnetic-field change, which cause sufficient transient losses to rapidly transitioning the magnet conductor to the normal state, and the transfer of part of the magnet's stored energy from the magnet coils to the ESC coils.
	
	ESC combined ability of fast quench initiation and energy extraction allows achieving excellent quench protection performance, i.e. low hot-spot temperature and peak voltage to ground.
	Furthermore, ESC does not require direct electrical connection to the magnet circuit, nor does it rely on thermal diffusion across thin insulation layers.
	When including ESC in a magnet design, interaction between ESC and other sub-systems must be carefully analyzed, as well as the electromagnetic forces acting on the magnet and ESC coils.
	
	The case study of the protection of a 15~m long, 12~T racetrack magnet using this new method is described.
	The electro-magnetic and thermal transients occurring in the magnet during the discharge are simulated with the STEAM-LEDET program.
	ESC quench protection performance is compared to those of two conventional protection methods, namely quench heaters and CLIQ, and shows remarkable improvement.
	
	ESC technology can be integrated into the design phase of future superconducting magnets.
	It is envisaged that less stabilizer will be needed to meet quench protection requirements when ESC is introduced, leading to more efficient, compact, and cost-effective magnet designs.

	\clearpage

	\bibliography{IEEEabrv,referencesIEEE}
	\bibliographystyle{unsrt}
	
\end{document}